\documentclass[twocolumn,showpacs,amsmath,amssymb,superscriptaddress,aps,pre]{revtex4-1}
\pdfoutput=1
\usepackage{graphicx,color} 
\usepackage{dcolumn}
\usepackage{bm,bbold,empheq,mathdots}
\usepackage{units,enumitem}
\usepackage{hyperref}

\def\mydoubleq#1{``#1''}
\newcommand{\ba}{\begin{eqnarray}}
\newcommand{\ea}{\end{eqnarray}}
\newcommand{\be}{\begin{equation}}
\newcommand{\ee}{\end{equation}}

\begin{document}

\title{Bundling in brushes of directed and semiflexible  polymers}
\author{Panayotis Benetatos}  \email{pben@knu.ac.kr}  \affiliation{Department of Physics, Kyungpook National University, 80 Daehakro, Bukgu, Daegu 702-701, Korea}
\author{Eugene M. Terentjev} \affiliation{Cavendish Laboratory, University of Cambridge, 19 J.~J.~Thomson
  Avenue, Cambridge, CB3 0HE, United Kingdom}

\author{Annette Zippelius} \affiliation{Institute for Theoretical
  Physics, Georg-August-Universit\"at G\"ottingen,
  Friedrich-Hund-Platz 1, 37077 G\"ottingen, Germany} \affiliation{Max
  Planck Institute for Dynamics \& Self-Organization, Am Fassberg 17, 37077 G\"ottingen, Germany}

\date{\today}  

\begin{abstract}
  We explore the effect of an attractive interaction between parallel
  aligned polymers, which are perpendicularly grafted on a
  substrate. Such an attractive interaction could be due to, e.g.,
  reversible cross-links. The competition between permanent grafting
  favoring a homogeneous state of the polymer brush and the attraction, which tends to
  induce in-plane collapse of the aligned polymers, gives rise to an
  instability of the homogeneous phase to a bundled state. In this
  latter state the in-plane translational symmetry is spontaneously
  broken and the density is modulated with a finite wavelength, which
  is set by the length scale of transverse fluctuations of the grafted
  polymers.  We analyze the instability for two models of aligned
  polymers: directed polymers with line tension $\epsilon$ and weakly
  bending chains with bending stiffness $\kappa$.
\end{abstract}
\maketitle

\section{Introduction}

Many important biopolymers, such as the structural elements of the cytoskeleton (F-actin, microtubules, intermediate filaments) exhibit semiflexibility, that is,  behavior intermediate between that of a random coil and a slender rod \cite{Howard}. Their mechanical and thermodynamic properties are dominated by their bending rigidity which gives rise to a finite persistence length defined as the correlation length of their straightness along their contour. Many of these polymers form bundles of many parallel-aligned and cross-linked filaments \cite{Bartles,Ayscough}. These bundles have been the the subject of intense study in recent years, because they form the structural components of various cellular processes such as stereocilia, filopodia, and microvilli. Experiments on {\it in vitro} reconstituted actin bundles have revealed the important role of reversible cross-linking by actin-binding proteins (ABPs)\cite{Lieleg,Gardel,Lieleg_Bausch,Sackmann}. Monte Carlo simulations of supramolecular self-assembly based on a model of patchy hard spheres have displayed a bundling transition in an interacting polymer gas \cite{Fasolino}.  Theoretical attempts to describe bundling of semiflexible polymers as a thermodynamic phase transition, effectively reduce it to the thermal binding-unbinding transition of two weakly bending (because of their large persistence length) semiflexible polymers\cite{Kierf_bundle,Heus_Vink}. This transition is driven by the interplay of entropy associated with the transverse undulations of the filaments and the potential energy of their attractive interaction (which can be due to reversible cross-linking)\cite{PB_depin,Kierf_depin}. The thermodynamic limit associated with this transition is in the direction of the polymer contour. When bundling involves more than two filaments, the relevant thermodynamic limit would involve an infinite array of parallel-aligned interacting filaments with constant areal density (in the plane perpendicular to the direction of the polymer contour). 

The effect of reversible cross-links on an infinite array of parallel-aligned filaments is going to depend on the boundary conditions imposed on the filament end points. If the filaments are grafted on a planar substrate, then we have a polymer brush. In recent years, there has been a growing interest in cross-linked polymer brushes because of promising technological applications \cite{Ballauff,Loveless}.  Cross-link induced brush collapse has been proposed as a mechanism for the selective gating in the nuclear pore complex which regulates cargo transport between the cytoplasm and the nucleus in eukaryotic  cells \cite{Lim}.   Grafted arrays of semiflexible chains in a poor solvent have been studied using Monte Carlo simulations in \cite{PPP}. In this case, the attractive interaction between polymers comes from the unfavorable polymer-solvent enthalpic interaction which causes minimization of their surface area.  In \cite{BW1}, scaling arguments concerning a similar system predict towers (bundles) or toroidal micelles. A uniformly  tilted phase is predicted in \cite{BW2} using an effective two-chain model proposed for moderately poor solvents involving not too high monomer concentration. Tilted phases of surfactants on surfaces modeled as hard rods grafted on a lattice have been analyzed in \cite{Safran}. Experimental evidence for the formation of bundles as a result of the interplay between permanent grafting and attractive interaction has been provided in experiments with vertically grafted carbon nanotube forests subjected to attractive capillary forces \cite{Lau}. Elastocapillary coalescence of bundles in a brush which is withdrawn from a perfectly wetting liquid has been studied in \cite{Boudaoud}.

\begin{figure}
\includegraphics[
width=0.4\textwidth]{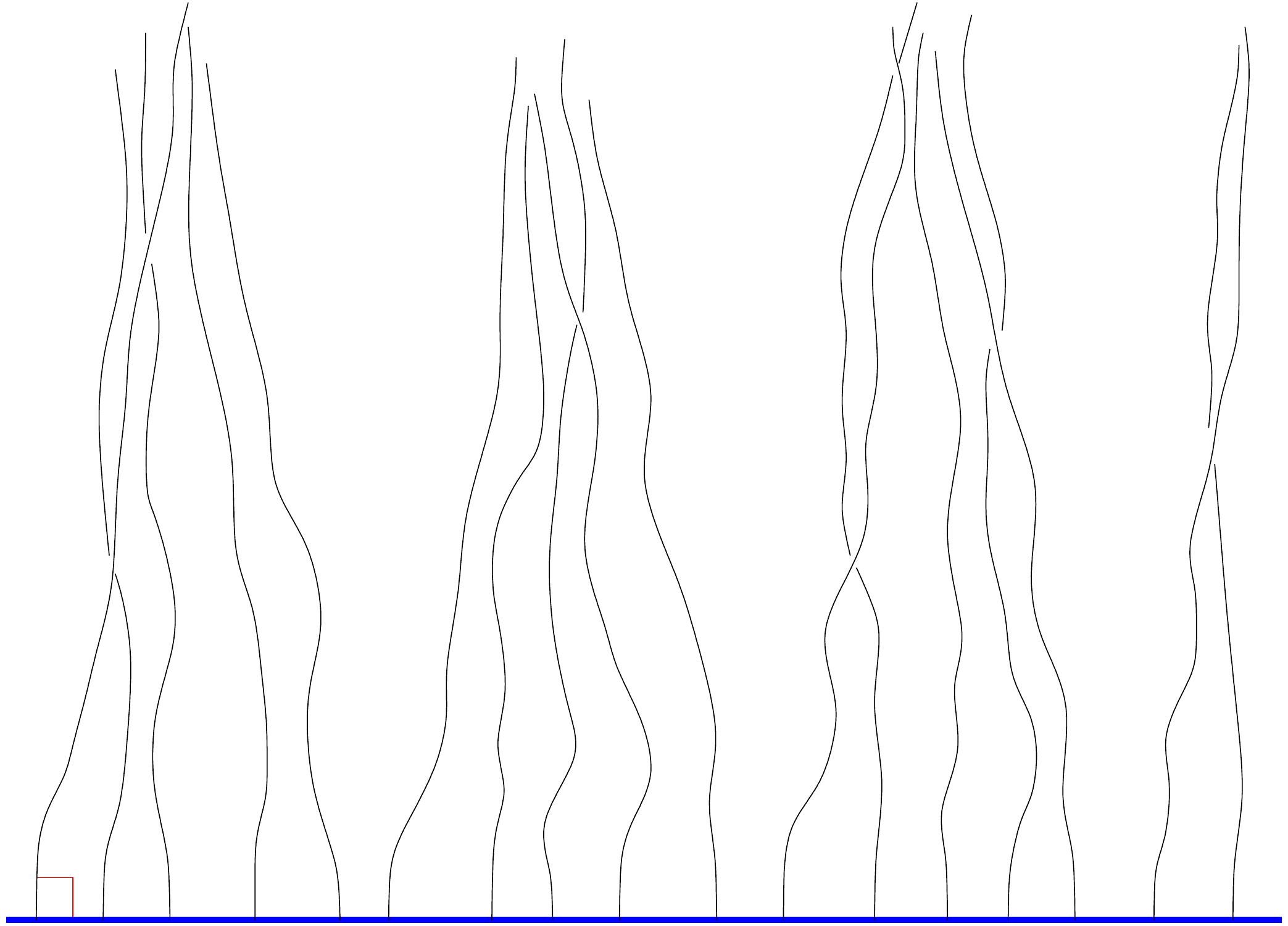}\caption{(Color online) A schematic diagram of a brush of weakly bending semiflexible polymers in the bundled state.
\label{bundled}}
\end{figure}

In this work, we analytically investigate the stability of a system of parallel-aligned, uniformly grafted vertical polymers to an attractive interaction. We propose a semimicroscopic model and treat  the random grafting positions on the planar substrate as quenched disorder. The polymers are held vertically either due to a tension (directed polymers) which penalizes tilting or due to their bending rigidity (semiflexible chains) which gives rise to a large persistence length. Both types of polymers give qualitatively similar results. As we increase the strength of the attractive interaction, both systems give rise to a bundled state with periodic modulation of the in-plane areal density. Such a state is schematically shown in Fig. ~\ref{bundled}. The order parameter for bundling is the Fourier transform of the average in-plane local areal density, analogous to the order parameter for crystallization. In the absence of any attractive interaction, the structure of the brush will reflect the disorder of the grafting points and the order parameter will vanish for every non-zero wavenumber. Bundling is signaled by the emergence of a non-vanishing order parameter at a finite wavenumber. In our model, each polymer is stretched individually (due to its tension or bending rigidity). However, in dense flexible brushes, stretching is usually the collective effect of mutual excluded volume interaction. Although we do not consider this effect in the present paper, our general conclusions are expected to hold generally because they are based on the interplay of fixed grafting and attractive interaction.

The paper is organized as follows: In Section II we introduce the model of directed polymers with tilting elasticity and an attractive interaction. In Section III, we present a coarse-graining procedure which leads to a mean-field free energy as a function of the order parameter. The instability of the disordered state resulting from the attractive interaction is analyzed in Section IV. The effect of an excluded-volume interaction is discussed in Section V. In Section VI, we go through the same analysis for a system of weakly bending semiflexible polymers. We conclude and discuss further extensions of this work in Section VI.

\section{Model} 

\begin{figure}
\includegraphics[
width=0.4\textwidth]{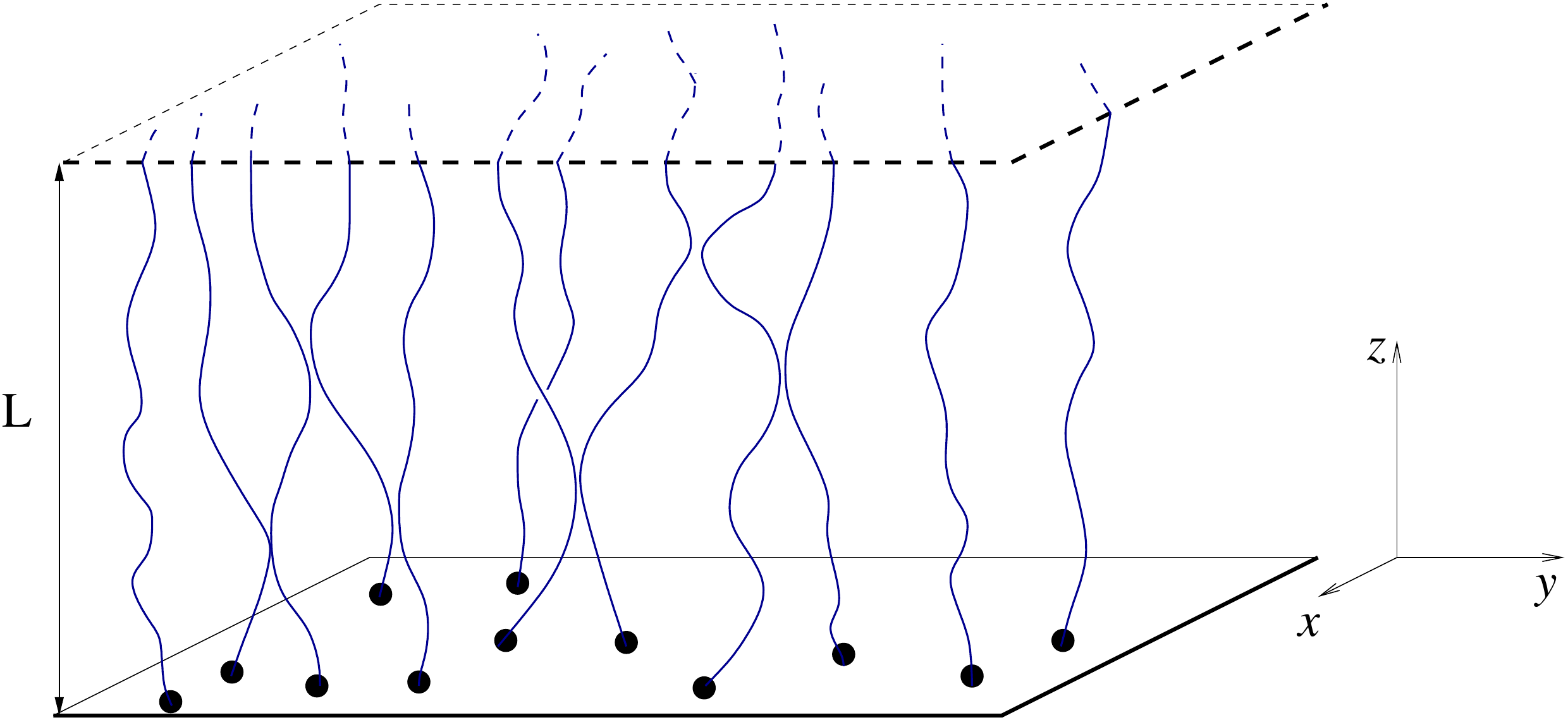}\caption{(Color online) A schematic diagram of a brush of directed
  polymers of thickness (height) $L$. $z$ is the preferred direction and we refer to $x,y$ as the \emph{transverse} or \emph{in-plane} direction.
\label{brush_schematic}}
\end{figure}

We consider $N$ directed polymers permanently grafted on a flat
surface. The stretching direction of the polymers is perpendicular to
the grafting surface and their free end points are free to slide
on a plane at a distance $L$ from the surface. Each polymer configuration is described by a curve
(path) ${\bf r}(z)=\bigl( x(z),y(z) \bigr)$, where $z \in [0,L]$ and
$z$ is the direction of alignment (Fig.~\ref{brush_schematic}). By the
definition of directedness, these paths exclude loops and
overhangs. The areal density  of the system in the $xy$ plane is
$\sigma=N/A$. The conformation of polymer $i$ is expressed as
\begin{equation}
\label{path}
{\bf r}_i (z) ={\bf R}_i+\int_0^z dz'{\bf t}_i(z')\;,
\end{equation}
where ${\bf R}_i$ is the position of the grafted end and ${\bf t}_i(z)\equiv {d
  {\bf r}_i(z)}/{dz}$ is the projection of the tangent vector on the $xy$-plane. The grafting points, ${\bf R}_i$,
are assumed to be uniformly distributed in the plane of the
surface. We treat this randomness as quenched disorder. We assume free boundary conditions at $z=L$, 
allowing the free polymer end to assume any arbitrary position on the
top plane with any slope. At the grafted end, the polymer has a fixed
position but it is free to fluctuate with any slope.

 The effective free-energy functional (``Hamiltonian'') of the directed
polymers consists of three terms:
\begin{equation}
\label{Ham1}
{\cal H} ={\cal H}_0+ {\cal H}_{ev}  + {\cal H}_{X} \;. 
\end{equation}
The first term 
\begin{equation}
{\cal H}_0=\frac{\epsilon}{2}\sum_{i=1}^{N}\int_0^L dz \,{\bf t}^2_i(z)
\end{equation}
 penalizes tilting away from the $z$ direction
with $\epsilon$ being the effective line tension \cite{DRN_book}. The second term accounts for excluded volume interactions and the third term
expresses the effective
attraction ($\mu>0$) due to reversible cross-linking
\begin{equation}
 {\cal H}_{X}  =-\frac{\mu}{2L}\int_0^L dz \sum_{i\neq j}V({\bf r}_i (z)-{\bf r}_j (z))\;.
\end{equation}
The function $V(r)$ is short ranged, intended to represent an effective interaction resulting from reversible cross-links \cite{Kierf_depin,Lipowsky}. Here it is  taken as a Gaussian with range $b$, $V(r)=\exp[-r^2/(2b^2)]$, assuming a harmonic interaction between the reversible cross-links~\cite{Broderix2002}.

\section{Free Energy}

The partition function of the system for a specific configuration of
grafting points, $\mathcal{C}\equiv\{{\bf R}_i\}$, reads 
\begin{eqnarray}
\label{Z_C}
Z(\mathcal{C})=\int \mathcal{D}\{{\bf t}_i\} e^{-{\cal H}_{ev}-{\cal H}_{X}},
\end{eqnarray}
where the measure  $\mathcal{D}\{{\bf t}_i\}$ denotes integration over
all possible polymer conformations with weight $\exp{(-{\cal H}_0})$ and consistent with $\mathcal{C}$. 
Here and in the following we set $k_{\rm B} T =1$. Physical
observables of interest can be calculated from the quenched-disorder
averaged free energy, $F=- [\ln Z]$, where $[...]$ denotes
average over all realizations of grafting points: 
\begin{eqnarray}
\label{quenched_ave}
[...]=\int d^2R_1d^2R_2...d^2R_N \frac{1}{A^N}(...) \; .
\end{eqnarray}

In this paper, we investigate the possibilty of a phase transition
from a state of homogeneous density to a density-modulated phase,
induced by the competition between the attractive interactions due to
reversible cross-linking and the excluded volume interactions (due to fixed grafting) favoring
the homogeneous state. As a first step in the calculation, we are
going to ignore the explicit excluded-volume interaction  (${\cal H}_{ev}$),
because the permanent homogeneous grafting is sufficient to prevent the in-plane
collapse of the polymers. In a second step, we will  investigate the
effects of excluded volume (Section V).

Going to Fourier space, we express $Z(\mathcal{C})$ as
\begin{equation}
\label{Z_C_Fourier}
Z(\mathcal{C})=\int \mathcal{D}\{{\bf t}_i\}
\exp\Bigg(- \frac{\mu N^2}{2LA}\int_0^L dz \sum_{\bf
    k}V({\bf k})|\rho({\bf k}, z)|^2\Bigg),
\end{equation}
in terms of the Fourier transform of the local areal density
\begin{eqnarray}
\label{Q}
\rho({\bf k}, z)=\frac{1}{N}\displaystyle\sum_{i=1}^N \exp\big(i {\bf k}\cdot
{\bf r}_i(z)\big)\;,
\end{eqnarray}
where the two-dimensional wave vectors ${\bf k}$ are consistent with
periodic boundary conditions in the $x$ and $y$ directions. We effectively decouple the
polymers using a Hubbard-Stratonovich transformation which introduces
the collective field $\Omega({\bf k},z)$:
\begin{equation}
\label{H_S}
Z(\mathcal{C})=\int \mathcal{D}\{\Omega\} e^{-Nf(\{\Omega\})},
\end{equation}
with the Landau-Wilson type free energy per polymer
\begin{align}
\label{free_energy}
f(\{\Omega\})&= \frac{\mu\sigma }{L}\int_0^L dz 
\sum_{{\bf k}\cdot {\bf
    n}>0}V({\bf k})|\Omega({\bf k}, z)|^2 \nonumber\\
&-\frac{1}{N}\sum_{i=1}^{N}\ln \mathfrak{z}({\bf R}_i)
\end{align}
The single-polymer partition function is given by 
\begin{align}
\label{single_polymer_partition}
\nonumber  \mathfrak{z}({\bf R}_i)= &\int \mathcal{D}\{{\bf t}_i\} \\
\times &\exp{\Big(\frac{\mu\sigma}{L}\int_0^L dz \sum_{{\bf k}\cdot{\bf n}>0}{\rm Re} (V({\bf k})\Omega({\bf k},z) e^{-i{\bf
  k}\cdot {\bf r}_i(z)})\Big)}.
\end{align}
The restriction ${\bf k}\cdot {\bf
    n}>0$ with ${\bf n}$ a two-dimensional unit vector has been
  introduced to avoid double counting and the integration measure is 
\begin{align}
\label{DOmega}
\mathcal{D}\{\Omega\}\equiv\prod_{{\bf k}\cdot {\bf
    n}>0} \frac{\mu N^2}{\pi A} d({\rm Re}( \Omega({\bf k},z)))d({\rm Im}( \Omega({\bf k},z)))\;.
\end{align}

In this paper, we are only going to discuss the saddle-point
approximation to the free energy Eq.~(\ref{free_energy}), i.e., we replace
$\Omega({\bf k},z)$ by $\Omega^{sp}({\bf k},z)$, which makes the free
energy Eq.~(\ref{free_energy}) stationary. The disorder averaged free energy reads in the saddle-point approximation
\begin{align}
\label{spa_free_energy}
f^{sp}&=[f(\{\Omega^{sp}\})]=-[\ln \mathfrak{z}({\bf R}_i)]\nonumber\\
&+\frac{\mu\sigma }{L}\int_0^L dz 
\sum_{{\bf k}\cdot {\bf
    n}>0}V({\bf k}) |\Omega^{sp}({\bf k}, z)|^2.
\end{align}
The above free energy is similar to model K of
Fredrickson~\cite{Fredrickson2006}. From the above equation, it is
apparent that correlations between the grafting points, $\{{\bf
  R}_i\}$, are irrelevant in the saddle-point approximation.

The order parameter suggested by the Hubbard-Stratonovich decoupling
is
\begin{align}
\label{order_par}
\Omega({\bf k},z)=\frac{1}{N}\sum_{i=1}^N\exp\big(i{\bf k}\cdot({\bf R}_i+\int_0^z dz'{\bf t}_i(z'))\big)\,.
\end{align}
This is the Fourier transform of the in-plane local areal density of the
directed polymers at the level $z$. In the absence of interactions,
its average value (averaging over both thermal fluctuations and
quenched disorder) is liquid-like (disordered): it vanishes identically for any
nonzero ${\bf k}$. The reason is the random, uncorrelated distribution
of the grafting positions ${\bf R}_i$. 

The saddle-point order
parameter obtained from the stationarity condition for the free energy
of Eq.~(\ref{spa_free_energy}), satisfies the
self-consistent equation
\begin{align}
\label{order_par_sp}
\Omega({\bf k},z)=\frac{1}{N}\Big[\langle\sum_{i=1}^N\exp(i{\bf k}\cdot{\bf r}_i(z))\rangle\Big]\,,
\end{align}
where $\langle...\rangle$ denotes average with the weight of the
partition function $\mathfrak{z}$ of Eq.~(\ref{single_polymer_partition}). An attractive
interaction will tend to cause in-plane collapse of the polymers. This
collapse is prevented by their permanent grafting on the surface. The
interplay of attractive interaction on the one hand and grafting and
tilting rigidity on the other may give rise to phases with periodicity
in the areal density (away from the plane $z=0$). Such a phase would
be described by a finite order parameter at a wave vector ${\bf k}\neq0$.


\section{ Instability of the disordered state}

To address the stability of the disordered phase,
we expand $[\ln \mathfrak{z}]$ up to quadratic order in
$\Omega^{sp}$ \cite{Fredrickson1992}:
\begin{align}
\label{quadratic_free_energy}
f^{sp}&=[f(\{\Omega^{sp}\})]\nonumber\\
&= \frac{\mu\sigma }{L}\int_0^L dz_1dz_2 
\sum_{{\bf k}\cdot {\bf
    n}>0}V({\bf k})\Omega^{sp}({\bf k}, z_1) \Omega^{sp}(-{\bf k}, z_2)\nonumber\\
\times&\Big(\delta(z_1-z_2)- \frac{\mu\sigma}{4L} V({-\bf k})A_k(z_1,z_2)\Big)
\end{align}
with the connected one-polymer correlation function
\begin{align}
&A_k(z_1,z_2)=\nonumber\\
&[\langle e^{i{\bf k}\cdot ({\bf r}_i(z_1)-{\bf r}_i(z_2))}\rangle_0-\langle e^{i{\bf k}\cdot {\bf r}_i(z_1)}\rangle_0 \langle e^{-i{\bf k} \cdot {\bf r}_i(z_2}\rangle_0]\nonumber
\end{align}
evaluated with ${\mathcal H}_0$.

To assess the full stability of the disordered state, one needs to
diagonalise $A_k(z_1,z_2)$ which is difficult in general. As a first
step, we replace the $z$-dependent order parameter, $\Omega({\bf k},
z)$, by its average along $z$: $\Omega({\bf k})=(1/L) \int_0^L dz \Omega({\bf k},z).$ This $z$-independent
approximation for the order parameter is expected to hold for
well-entangled (dense) brushes with an entanglement length $\;l_z \ll L$. The
entanglement length $l_z$ (also called deflection length) is defined as the distance along the $z$ axis
needed for the polymer to wander before it collides with its nearest
neighbor \cite{DRN_book}: \be \langle |{\bf r}_i(l_z)-{\bf
  r}_i(0)|^2\rangle=\frac{A}{N}\;, \ee which implies \be
l_z=\frac{\epsilon }{2  }\frac{A}{N} \;. \ee

\begin{figure}
\includegraphics[
width=0.4\textwidth]{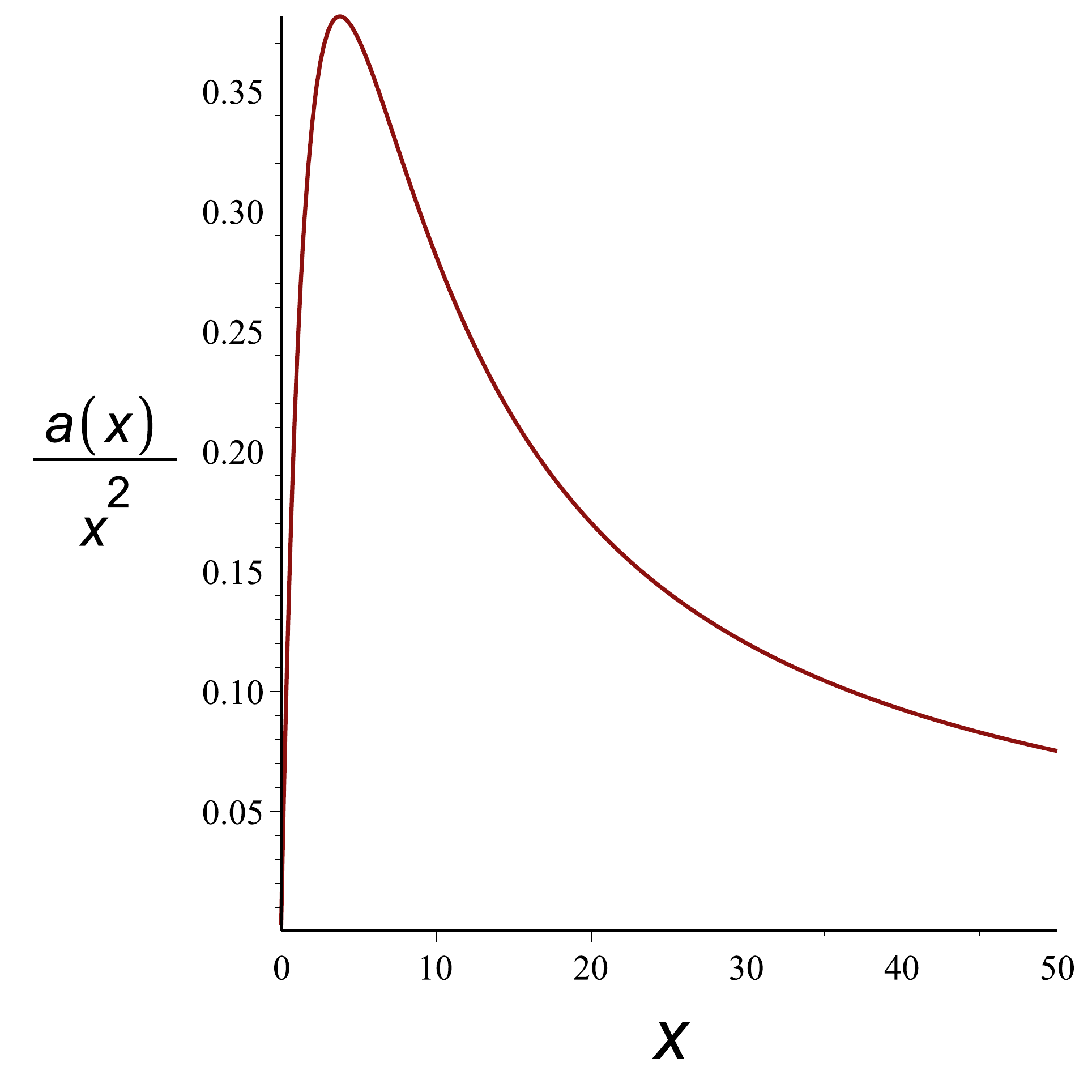}\caption{ Connected correlation 
$a(x)/x^2$, whose maximum determines the wavenumber of the mode which first becomes unstable for a brush of directed polymers 
  \label{connected_correlation}}
\end{figure}

Given this approximation, the coefficient of the quadratic term in
Eq.~(\ref{quadratic_free_energy}) simplifies considerably and
implies stability (associated with a positive sign) provided
\begin{equation}
\label{stability}
1>\frac{\mu\sigma}{4}(\frac{\epsilon}{Lk^2})^2V(k)a(\frac{Lk^2}{\epsilon})\;,
\end{equation}
where $a(x)\equiv 4x-12+16e^{-x/2}-4e^{-x}$.
An instability occurs, when the above inequality is violated. Whether or
not this happens at  a finite wavenumber $k$, depends on the function
$a(x)/x^2$, which is
is plotted in Fig.~\ref{connected_correlation}. The coefficient
becomes simpler in the two extreme cases of the range of the attractive
potential. In the short-range limit, which is also the most realistic,
the range $b$ of the attractive interaction is much smaller than the
in-plane radius of gyration of a free (non-interacting) polymer: $b \ll
\sqrt{L/\epsilon}$. In this case, the mode of the local (in the
$xy$-plane) areal density which corresponds to the peak of $\epsilon^2
a(L k^2/\epsilon)/(L k^2)^2$ at a wavenumber of the order of
$\sqrt{\epsilon /L}$ is the first to become unstable as the
dimensionless control parameter $\sigma \mu V(k\approx 0)$ grows
above a number of the order of one. It is this instability which
signals the formation of bundles which modulate the areal density at
the wavelength of the unstable mode. In the other extreme limit, the
range of the attractive interaction is much larger than the in-plane
radius of gyration of a free polymer: $b \gg \sqrt{L/\epsilon}$. In
that case, the instability occurs at a wavelength of the order of the
interaction range $b$ as $ \sigma \mu V(k\approx b^{-1})$ becomes greater than
a number of the order of one (assuming that the excluded volume
interaction is negligible: $\mu b^2/L \gg \lambda$). This limit of negligible radius of gyration compared to the range of the interaction would apply to the $T=0$ instabilities in attractive brushes such as those observed in Refs. \cite{Lau,Boudaoud}.

We point out that the emergence of an instability in the local ($xy$)
areal density at a finite wavelength is a robust result, not affected
by the $z$-independent approximation for the order parameter
$\Omega({\bf k}, z)$.  Indeed if we consider a $z$-dependent trial
function of the form \be
\label{sines}
\Omega ({\bf k},z)=\sum_{l=1}^{\infty} \omega({\bf k}, l)\sin\big(\frac{(2l-1)\pi}{2L} z\big)\;,
\ee
one can easily check that the diagonal elements of lowest order (in $l$) become negative first at a lower $k$.

The hallmark of bundling is the peak at a finite wavenumber of the
function $\epsilon a(L k^2/\epsilon)/(L^2 k^2)$. In order to gain
insight into the cause of this peak, we now investigate the stability
of a \mydoubleq{gas} of directed polymers with their end-points free to slide on
both planes (i.e., we remove the fixed-grafting constraints at the
bottom plane). In that case, there is no quenched disorder in the
system, and the disorder in the positions of the end-pints at $z=0$,
${\bf R}_i$, should be treated on an equal footing as the fluctuations in the
conformations of the slope [$\dot{\bf r}_i(z)$]. This is reflected in
the form of the saddle-point free energy per chain:
\begin{align}
\label{quadratic_free_energy_annealed}
f^{sp}&=\frac{\mu\sigma }{L}\int_0^L dz_1dz_2 
\sum_{{\bf k}\cdot {\bf
    n}>0}V({\bf k})\Omega^{sp}({\bf k}, z_1) \Omega^{sp}(-{\bf k}, z_2)\nonumber\\
&\Big(\delta(z_1-z_2)- \frac{\mu\sigma}{4L} V({-\bf k})B_k(z_1,z_2)\Big)
\end{align}
where now the one-polymer correlation function does not have a disconnected part
\begin{align}
B_k(z_1,z_2)=
\langle e^{i{\bf k}\cdot({\bf r}_i(z_1)-{\bf r}_i(z_2))}\rangle_0=e^{-k^2|z_1-z_2|/2\epsilon}
\end{align}
The function $B_k(z_1,z_2)$ is monotonically decaying as a function of the wavenumber $k$.
Hence, as we increase the strength of the attractive interaction, an
instability occurs at $k=0$ which corresponds to macroscopic phase
separation instead of bundling at a finite wavelength. We conclude that the peak in Fig. \ref{connected_correlation} which implies the formation of finite-size bundles is due to the quenched constraints of the grafting points.

\section{Effects of excluded volume}

So far, we have ignored the excluded volume interaction, arguing that
grafting stabilises our system against collapse. Now we show that
adding  the excluded volume interaction does not change the
results qualitatively, but contributes to an increase in the critical
attractive strength as one would expect.

We choose the excluded volume interaction \cite{Doi-Edwards} as
\begin{equation}
 {\cal H}_{ev}=\frac{\lambda}{2}\sum_{i\neq j}\int_0^L\delta({\bf r}_i(z)-{\bf r}_j(z)).
\end{equation}
Equation (\ref{Z_C_Fourier}) then reads 
\begin{equation}
\label{Z_C_Fourier_ev}
Z(\mathcal{C})=\int \mathcal{D}\{{\bf c}_i\}
\exp\Bigg(-\frac{N^2}{2A}\int_0^L dz \sum_{\bf
    k}\Lambda({\bf k})|\rho({\bf k}, z)|^2\Bigg)
\end{equation}
with $\Lambda({\bf k})=\lambda-\mu V(k)/L $.
The additional complication is the change of sign of $\Lambda(k)$ at a  wavenumber $k_0$ with  $\Lambda(k_0)=0$. 
This requires two Hubbard-Stratonovich transformations, one for $k<k_0$ and one for $k>k_0$. Apart from that, the calculation proceeds in complete analogy, so that we find for the saddle point free energy in quadratic order:
\begin{align}
\label{F_quadratic1}
\nonumber  f^{sp}&=\frac{\sigma L}{2}  \sum_{0\neq |{\bf
      k}|<k_0}|\Lambda(k)|\Big(1-|\Lambda(k)|
  \frac{\sigma{\epsilon}^2}{L\;k^4}\;a(\frac{L
    k^2}{\epsilon})\Big)|\Omega({\bf k})|^2\nonumber\\  
 &+\frac{\sigma L}{2} \sum_{|{\bf
      k}|>k_0}|\Lambda(k)|\Big(1+
  |\Lambda(k)|\frac{\sigma{\epsilon}^2}{L\;k^4}\;a(\frac{L
    k^2}{\epsilon})\Big)|\Omega({\bf k})|^2 \;,
\end{align}
The instability occurs, when
\begin{equation}
1=|\Lambda(k)|
  \frac{\sigma{\epsilon}^2}{L\;k^4}\;a(\frac{L
    k^2}{\epsilon}).
\end{equation}
Comparison with Eq. (\ref{stability}) reveals that the attractive
interaction is thus effectively reduced by the excluded volume
according to $\mu V(k)\to \mu V(k)-\lambda L$. The bundling transition
requires a correspondingly larger attractive interaction.

\section{Brushes of weakly bending semiflexible polymers}

\begin{figure}
\includegraphics[
width=0.4\textwidth]{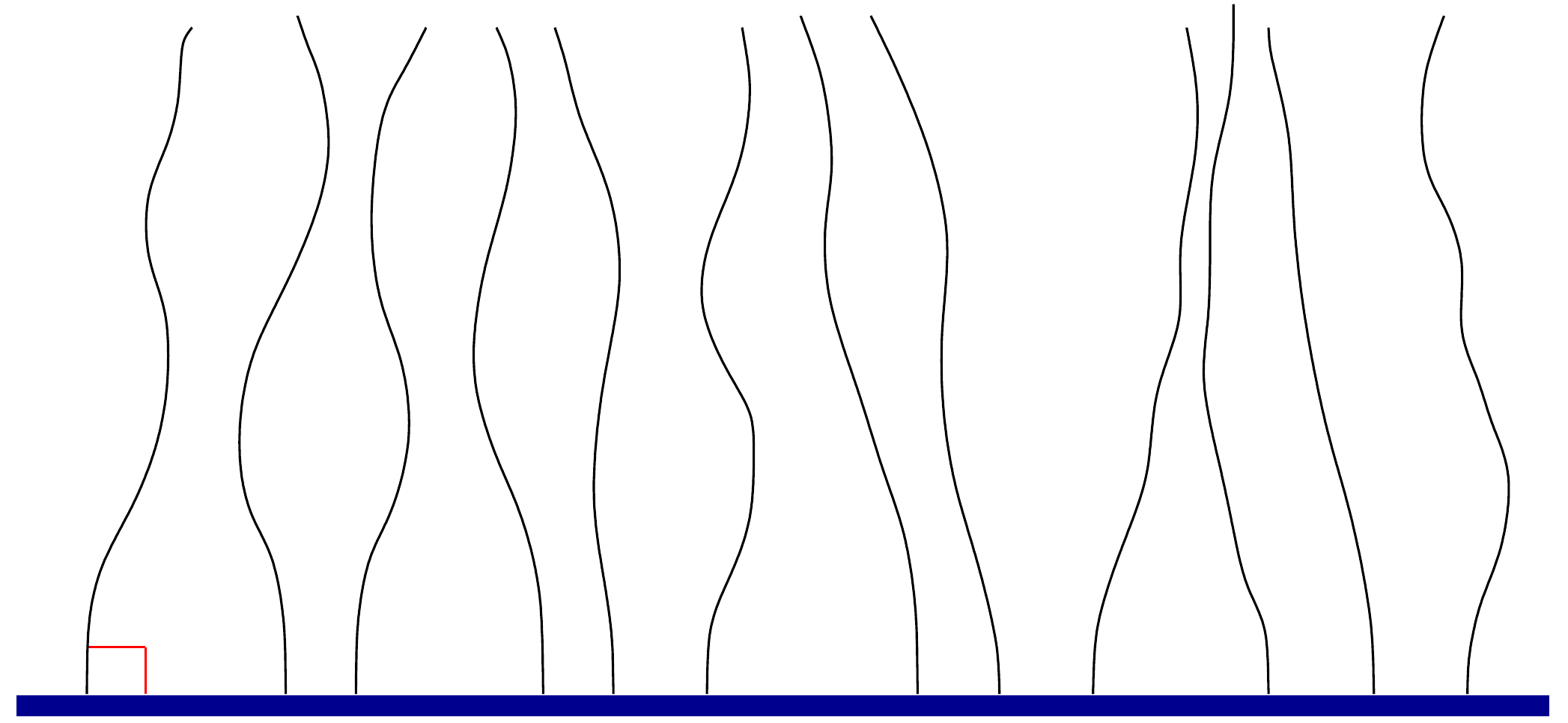}\caption{(Color online) A schematic diagram of a forest (brush) of weakly bending semiflexible
  polymers.
\label{forest_schematic}}
\end{figure}

The analysis and the qualitative results that we presented in the
preceding sections for a system of directed (flexible) polymers, carry
over for a system of identical perpendicularly grafted weakly bending
semiflexible polymers as shown schematically in
Fig. (\ref{forest_schematic}). The grafting positions are assumed
fixed and random, as in the case of directed polymers. We model a
semiflexible polymer of contour length $L$ as a locally inextensible fluctuating line with
bending rigidity. The weakly bending approximation holds when the
angle that the tangent vector along the polymer contour makes with the
grafting direction (in our case, $z$) is small. In this approximation,
we neglect fluctuations of polymer segments in the $z$-direction
compared to fluctuations in the transverse $xy$-plane.
 We parametrize
the conformation of polymer $i$ as in Eq. (\ref{path}).

The only difference to the directed polymer case, is the elastic energy
\begin{equation}
{\cal H}_0 \{{\bf r}_i(z)\}
=\frac{\kappa}{2}\sum_{i=1}^{N}\int_0^L dz \left(\frac{d{\bf t}_i(z)}{dz}\right)^{\!2},
\end{equation}
where now $\kappa$ is the bending rigidity, which penalises bending instead of tilting. It is related to the persistence length which is the
correlation length of the polymer's directedness $L_p$ via
$L_p=\kappa/(k_B T)$. We assume $L_p \gg L$ for the weakly bending
approximation to hold.
The excluded volume and the effective attractive interaction are taken
to be the same as in the case of directed polymers.  of the tangent
vector.

\begin{figure}
\includegraphics[
width=0.4\textwidth]{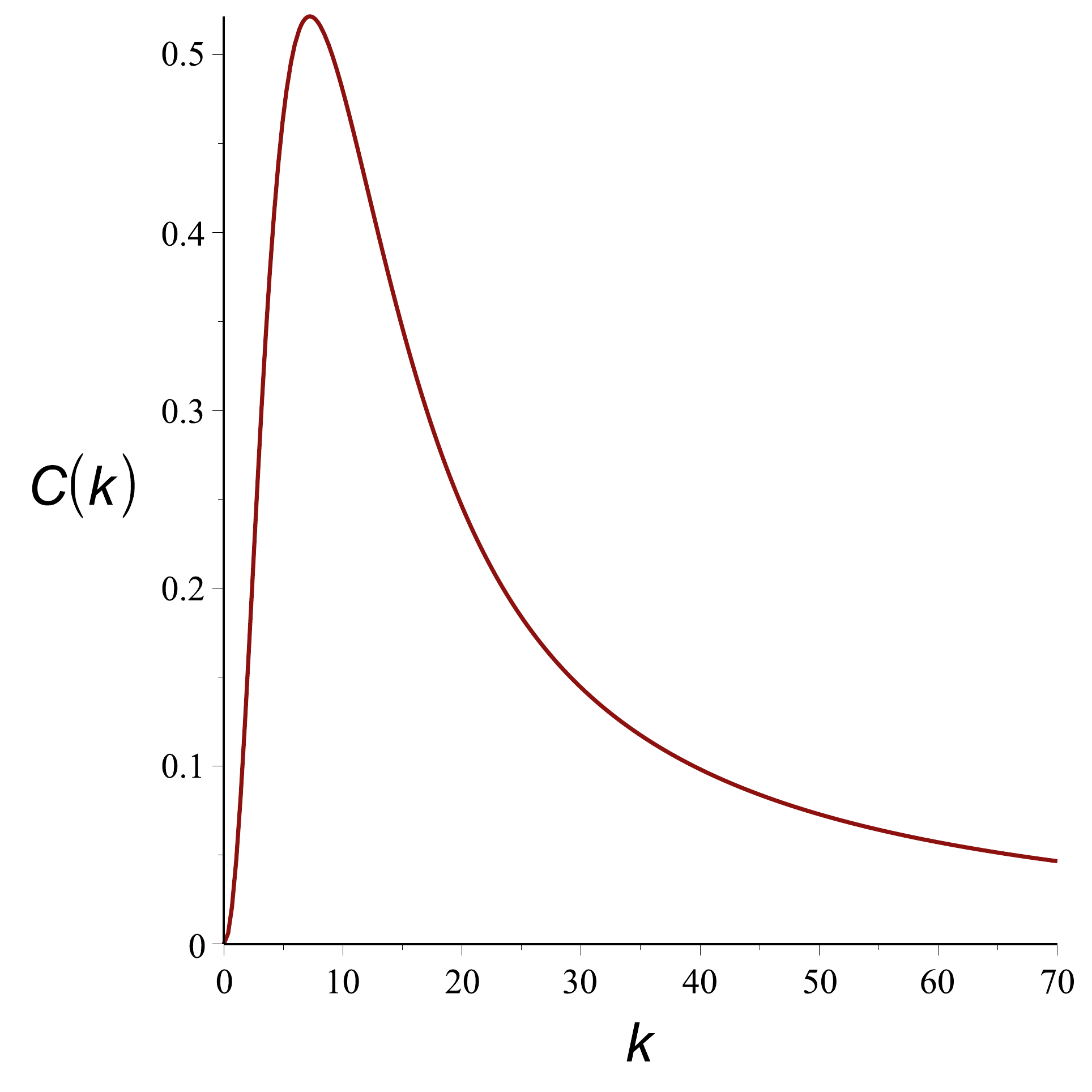}\caption{(Color online)  Connected correlation function
$C(k)$, whose maximum determines the wavenumber of the mode which first  becomes unstable in a brush of weakly bending semiflexible polymers. $k$ is measured in units of $L$ and, in this plot, $L/L_p=0.1$.
\label{c(k)}}
\end{figure}

Following similar steps as those which led to Eq. (\ref{F_quadratic1}), we obtain the quadratic part of the free energy:
\begin{align}  
 f^{sp}&=\frac{\sigma L}{2} \sum_{0\neq |{\bf
      k}|<k_c}|\Lambda( k)|\Big(1-\sigma
  |\Lambda(k)| L C(\frac{L^3
    k^2}{\kappa})\Big)|\Omega({\bf k})|^2 \nonumber\\
& +\frac{\sigma L}{2} \sum_{|{\bf
      k}|>k_c}|\Lambda(k)|\Big(1+\sigma
  |\Lambda(k)| L C(\frac{L^3
    k^2}{\kappa})\Big)|\Omega({\bf k})|^2\;,\nonumber
\end{align}
where 
\begin{align}
\label{C}
&C\big(\frac{L^3k^2}{\kappa}\big)=\\ \nonumber
&\int_0^1dz\int_0^1dz'\exp\Big(-\frac{L^3 k^2}{\kappa}\big(\frac{z^3}{6}+\frac{z'^3}{2}-\frac{z z'^2}{2}\big)\Big)\theta(z-z')\\ \nonumber
&+\int_0^1dz\int_0^1dz'\exp\Big(-\frac{L^3 k^2}{\kappa}\big(\frac{z'^3}{6}+\frac{z^3}{2}-\frac{z' z^2}{2}\big)\Big)\theta(z'-z)\\ \nonumber
&-\Big(\int_0^1 dz \exp\big(-\frac{L^3 k^2}{\kappa}z^3\big)\Big)^2 \;,
\end{align}
where $\theta(z)$ is the Heaviside step function. This is the characteristic function which replaces $a(x)/x^2$ in Eq. (\ref{F_quadratic1}).  This integral can be evaluated numerically. In Fig. (\ref{c(k)}), we plot $C(L^3 k^2/\kappa)$ as a function of $k$, where $k$ is measured in units of $L$, for $L/L_p=0.1$.

The results are very similar to those obtained for a brush of directed polymers, except
for the lengthscale, $L_{\perp}$, of transverse fluctuations of the free polymer end which gives the wavelength of the emerging periodic structure for short-range attractive interaction. For a
weakly bending chain, its scaling with the contour length is given by
$L_{\perp}=\sqrt{L^3/(3L_p)}$ \cite{Gholami2006}. For the relevant
case of short-range attractive interactions, the wavelength of the
bundled state is set by $L_{\perp}$ and the transition occurs when
$\sigma \mu V(k\approx 0)={\cal O}(1)$.

\section{Discussion and outlook}

In this paper, we have shown that the competing tendencies of the attractive interaction to bring the polymers closer and the fixed grafting points to keep them in place give rise to a modulated phase with periodicity in the in-plane areal density. We considered two types of parallel-aligned chains, namely, directed polymers with tilt stiffness and semiflexible polymers with bending stiffness. In both cases, we obtain qualitatively similar results. The former are easier to handle analytically and, when they capture the relevant universal behavior, they are very useful as structural elements of minimal theoretical models. They have already been used to investigate the gelation transition in a parallel-aligned array of directed polymers freely moving between two planes perpendicular to their preferred direction \cite{DPs1} and also in the stretching elasticity of a semiflexible polymer bundle with permanent cross-links \cite{NJP}. In the current system, for short range attractive interaction, the transition from a disordered (reflecting the disorder of the grafting points) to a periodic structure happens at a wavelength of the order of the in-plane radius of gyration of the free (non-interacting) chain.  In the case of attraction of longer range than the in-plane radius of gyration, the instability occurs at a wavelength corresponding to that range. The transition threshold depends on the strength of the attractive interaction, the areal density of the polymers, and the range of the interaction. An excluded-volume interaction increases the threshold for the instability, but it does not affect the critical wavelength. 

Our work can be extended in many interesting directions which will be presented in future publications. The leading-order (quadratic) term in the Landau-Wilson free energy which is kept in the previous analysis allows us to show the instability of the disordered state, but it does not give us the actual structure of the modulated phase (e.g., hexagonal lattice, striped phase, etc.). That would require keeping higher order terms in the free energy. In addition, the precise meaning of the critical wavelength that we obtain here (the mode which first becomes unstable in the quadratic free energy) depends on the type of transition. For a continuous transition, it would coincide with the wavelength of the emerging periodic structure at the transition point. For a first order transition, it would coincide with  with the wavelength of the emerging periodic structure at the point where the disordered metastable minimum disappears. The crystalline order is discussed in the context of mean field theory. Fluctuations are expected to yield algebraic correlations similar to those in the X-Y model \cite{KT}. In this paper, we considered a short-range  attractive interaction between the polymer chains, intended to express the effect of reversible cross-links. With the notable exception of \cite{perm_xlink_brush} which focuses on the percolation transition, the theoretical investigation of permanently cross-linked brushes remains open. Permanent cross-links are expected to induce an effective attraction; on the other hand, they also tend to freeze the disorder of the uncross-linked phase, and it is not clear which tendency wins in the end.

 \begin{acknowledgements}
Financial
support by the Deutsche Forschungsgemeinschaft through Grant
SFB-937/A1 is gratefully acknowlegded. P.B. acknowledges support by the Kyungpook National University Research Fund, 2012.
\end{acknowledgements}

\bibliography{brush_bundleNotes27}

\end{document}